\documentclass[aps,twocolumn,groupedaddress,showpacs,floatfix]{revtex4}

\usepackage{graphicx}

\begin{document}
\bibliographystyle{apsrev}

\title{Susceptibility of a spinon Fermi surface coupled to a U(1) gauge field}

\author{Cody P. Nave$^1$, Sung-Sik Lee$^2$ and Patrick A. Lee$^1$}
\affiliation{
$^{1}$Department of Physics, Massachusetts Institute of Technology,
Cambridge, Massachusetts 02139\\
$^{2}$Kavli Institute for Theoretical Physics, University of California,
Santa Barbara, California 93106, U.S.A.}
\date{\today}

\begin{abstract}
We study the theory of a $U(1)$ gauge field coupled to a spinon Fermi surface.
Recently this model has been proposed as a possible description of the organic compound
$\kappa$-(BEDT-TTF)$_2$Cu$_2$(CN)$_3$.  We calculate the susceptibility of 
this system and in particular examine the effect of pairing of the underlying
spin liquid.  We show that this proposed theory is consistent with the observed
susceptibility measurements.
\end{abstract}

\maketitle

The organic compounds $\kappa$-(BEDT-TTF)$_2$X are an interesting class
of materials.
Recent experiments have shown promise that this compound 
where the anion, X, is Cu$_2$(CN)$_3$ may be the first experimental realized spin liquid.  
This material can be described as a nearly isotropic effectively two
dimensional spin 1/2 triangular lattice at half-filling.  Experimentally, this
material is found to be insulating and yet it has no long range magnetic ordering 
down to mK temperatures. Also the static spin susceptibility remains finite 
down to the lowest temperatures measured.  \cite{Shimizu:03}
These observations have led to the proposal that the state
may be described by a spinon Fermi surface coupled to a $U(1)$ gauge
field. \cite{Motrunich:05,SSLee:05}
The susceptibility is fit with the high temperature series expansion
of the spin 1/2 Heisenberg model on a triangular lattice.  From this fit,
the exchange coupling $J$ is found to be around 250 K. In addition,
the susceptibility is found to drop sharply at low temperatures around
10 K before saturating to a finite value.\cite{Shimizu:03}

Recent measurements of the specific heat have suggested the existence of a peak
in the electronic specific heat at around 6 K, once the phonon contribution has been
subtracted away. \cite{Nakazawa:06}  Led by this discovery, it was proposed that the $U(1)$
 spin liquid state may have some sort of pairing instability.  \cite{SSLee:06}  
Since the specific heat was also found to be unaffected by a magnetic field of up to 8T, 
conventional singlet pairing is unlikely.  The pairing could, however, be ordinary BCS triplet 
pairing or a new kind of pairing.  Recently Lee \textit{et al.} \cite{SSLee:06} proposed 
a possible new kind of pairing called ``Amperean''pairing. 
Unlike normal BCS pairing across the Fermi surface, this pairing is 
between two spinons on the same side of the Fermi
surface.  In particular, in the Amperean paired state, one
pairs the spin with momentum $\mathbf{Q} + \mathbf{p}$ with the 
spin with momentum $\mathbf{Q} - \mathbf{p}$ where
$|\mathbf{Q}| = k_\mathrm{F}$ and $|\mathbf{p}|$ small. The Amperean pairing can occur 
between two particles carrying almost parallel momenta 
due to the attractive interaction mediated by the magnetic 
fluctuations of the emergent gauge field. 
As a result the pairs carry net momentum
$2 k_\mathrm{F}$ as opposed to 0 in the BCS state. In particular,
the authors showed that it is possible for there to be an instability to this kind of
pairing for the spinon Fermi surface coupled to a $U(1)$ gauge field.
They also derived a number of experimental consequences of this model and show how they could explain
many of the features seen in the actual experiments on $\kappa$-(BEDT-TTF)$_2$Cu$_2$(CN)$_3$
Here we calculate the effect of pairing on the zero-field spin susceptibility of 
such a system and compare the result to what is experimentally seen in 
this organic compound.

Starting from a spinon Fermi surface, it is clear that at $T=0$ the spinons
give rise to a Pauli paramagnetic term due to the non-zero density of
states.  Standard BCS singlet pairing, however, leads to the reduction of this paramagnetism
as a gap opens. At first sight, this seems to provide a natural explanation of the sharp
drop in susceptibility below 10 K. 
However we have already excluded BCS singlet pairing 
because it is inconsistent with the observed
insensitivity of the specific heat to magnetic field.
Both triplet BCS pairing and alternate types of pairing such as LOFF and Amperean
are consistent with the specific heat measurement.
However it turns out that for such pairing states,
the spinon contribution to the Pauli paramagnetism is 
unaffected by the onset of pairing, which seems
inconsistent with the observed drop of susceptibility at low temperatures.
In this paper, we will show that the drop of susceptibility can be explained 
if the effect of gauge fluctuations is taken into account.
Before we include the effect of gauge fluctuations, 
below we first ignore the gauge fluctuations and 
explain why the onset of pairing does not affect the contribution of spinons 
to the spin susceptibility in the Amperean, LOFF and triplet BCS pairing states.  

\begin{figure}
\centerline{\includegraphics[width=1.25in]{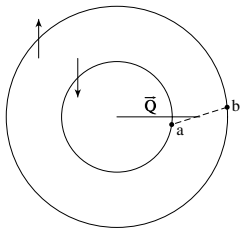}}
\caption{
In the Amperean pairing state, the spinons at a and b are paired.  
We see that by choosing $\Delta \mathbf{Q} + \mathbf{p}$ correctly, 
both a and b lie near their respective Fermi surfaces.
}
\label{pairing}
\end{figure}

To see this, we begin with a spinon system with a well defined Fermi surface.
Applying a magnetic field creates two different Fermi seas for the up and down spinons,
as shown in Fig. \ref{pairing}.  First we consider the case of Amperean pairing, where pairing
occurs on the same side of the Fermi surface.  It is possible for both of these spinons to 
lie near the Fermi surface even after the magnetic field has been applied (Fig. \ref{pairing}).
This is achieved by pairing the spin up spinon with momentum
$\mathbf{Q} + \Delta \mathbf{Q} + \mathbf{p}$ with the spin down spinon with
momentum $\mathbf{Q} - \Delta \mathbf{Q} - \mathbf{p}$, where $|\mathbf{p}| \ll k_\mathrm{F}$
and $ \Delta \mathbf{Q} = \left( \mu_\mathrm{B} H / v_\mathrm{F} \right)
\hat{\mathbf{Q}}$.
Moreover, the phase space available for $\mathbf{p}$ is unchanged with the applied field $H$,
as long as the curvature difference between spin up and down Fermi surfaces when $H \not= 0$
can be ignored.  Thus in this approximation, there is no Zeeman limiting field for this pairing.  Furthermore,
the susceptibility is not reduced by pairing because although the opening of the pairing
gap does smear out the momentum distribution $n_\mathbf{k}$, it leaves the occupied area of
up and down spins unchanged.  Thus the magnetization remains unchanged
despite the opening of the pairing gap, and 
the underlying spinon paramagnetism is not destroyed.  

We also note that this same argument holds for both the LOFF state
\cite{Fulde:64,Larkin:64} and the 
BCS triplet state.  In the LOFF state the pairing is between a spin up on 
the spin up Fermi surface and a spin down on the opposite side of the
spin down Fermi surface.  This state as well only smears
the momentum distribution leaving the occupied area unchanged and again
the underlying paramagnetism survives.  
In the particular case of BCS triplet pairing where only the equal spin pairings 
$\Delta_{\uparrow \uparrow}$ and $\Delta_{\downarrow \downarrow}$ are nonzero,
it is again clear that the pairing is unaffected by magnetic field and thus the
magnetization remains fixed despite the opening of a pairing gap.  In general
triplet pairing there is also a non-zero pairing between the up and down spins.
However, due to weak
spin-orbit coupling the spin quantization axis favors a particular
direction. When the applied field exceeds some small pinning field,
the quantization axis rotates in such a way that the equal spin
pairing description is appropriate and the thus the susceptibility remains
unchanged despite the pairing gap.

For completeness, we now show how this argument fails for the BCS singlet state.  
After splitting the spin up and spin down with a magnetic field, the paired spins 
do not lie on the Fermi surface.  If one insists on pairing $(\mathbf{k},\uparrow)$ 
with $(-\mathbf{k},\downarrow)$, the gap does not develop at the Fermi surface.
This is no longer the BCS singlet state, but in fact the breeched pairing state,
which is not energetically favorable for small $H$. \cite{Liu:03}

Thus we conclude that the naive application of triplet or Amperean pairing
does not explain the sharp drop in the spin susceptibility below 10 K.
In order to explain the data, we consider the effect of gauge fluctuations 
in the Amperean pairing state.  
We show that in the problem of a spinon Fermi surface coupled to a $U(1)$ gauge field, the
gauge field fluctuations give rise to a substantial paramagnetic contribution
to the susceptibility. 
The onset of Amperean pairing driven by the gauge fluctuations introduces a gap in the gauge
field by the Anderson-Higgs mechanism and suppress the gauge field contribution
to the susceptibility.  This suppression would account for the measured
drop in susceptibility that occurs around 10 K.  We note that this mechanism is 
independent of the form of pairing, whether it is triplet or Amperean.

We begin with the Lagrangian for a 2-D spinon Fermi surface system coupled to a $U(1)$ 
gauge field,
\begin{equation}
\mathcal{L} = \psi^\ast_\sigma \left( \partial_0 - i a_0 - \mu \right) \psi_\sigma +
\frac{1}{2 m}\psi^\ast_\sigma \left(-i \nabla - \mathbf{a}\right) \psi_\sigma.
\label{lagrangian}
\end{equation}
We have dropped the gauge field kinetic energy term because its strength is
inversely proportional to the charge gap. $\psi_\sigma$ is the spinon field and the
gauge field is $a = (a_0, \mathbf{a})$.  $\mu$ is the chemical potential.  We work
in Coulomb gauge $\nabla \cdot \mathbf{a} = 0$.

We begin by calculating the random-phase approximation (RPA) for this model.
The use of the RPA can be justified by the standard $1/N$ 
expansion. \cite{Ioffe:89,Lee:92,Polchinski:94}
We need to calculate the bare spinon polarization bubble shown in Fig. \ref{dressedpi} which
generates the gauge propagator.  Working with this bare spinon
bubble, i.e. not dressed by further fields inside the spinon loop, is 
equivalent to working to lowest order in the $1/N$ expansion.
The scalar i.e. longitudinal part of the gauge propagator is related to
the  density-density response and does not show singular behavior for small 
$\mathbf{q}$ and $\omega$.  
In other words the scalar part is screened out by spinon density fluctuations 
and we can focus on just the transverse part of the gauge field.  Because the gauge
field is now purely transverse, the spinon-gauge field vertex carries a vector index.

The gauge propagator is generated by a sum of spinon loops carrying a given spin.
We define the bare spinon polarization bubble to be
\begin{equation}
\Pi(q) = \sum_\sigma \Pi_\sigma(q),
\label{spinbubble}
\end{equation}
where
\begin{equation}
\Pi_\sigma(q) = \frac{1}{\beta} \sum_{k_0} 
\int \frac{d\, \mathbf{k}}{(2\pi)^2} \, G_\sigma \left(k+\frac{q}{2}\right) G_\sigma \left(k-\frac{q}{2}\right)
\left| \frac{ \mathbf{k} \times \hat{\mathbf{q}}}{m} \right|^2,
\end{equation}
with $q = (q_0, \mathbf{q})$ and $k$ similary defined.  We show the details
of this calculation in the appendix, but in the end we find that
\begin{equation}
\Pi_\sigma(q) = \frac{1}{2} \left( \frac{\gamma \, v^\sigma_\mathrm{F} \left|q_0\right|}
{\sqrt{\left(v^\sigma_\mathrm{F} \mathbf{q}\right)^2 + q_0^2} + \left| q_0 \right|} + 
\chi_\mathrm{d} \mathbf{q}^2 \right),
\label{dressedprop}
\end{equation}
where $\chi_\mathrm{d} = \frac{1}{12 \pi m}$ and $\gamma = \frac{k^\sigma_\mathrm{F}}{\pi}$.
$k^\sigma_\mathrm{F}$ and $v^\sigma_\mathrm{F}$ are the Fermi momentum and velocity
respectively for a spinon with spin $\sigma$ in an applied magnetic field.  We will
see that we need to keep the curvature of the Fermi surface so we use that
$k^\sigma_\mathrm{F} = \sqrt{k^2_\mathrm{F} \pm 2 m \mu_B H}$, where
$\sigma = (+,-)$ for the up and down spins respectively.

In order to calculate the susceptibility, we calculate the free energy of
the Lagrangian from Eq. \ref{lagrangian}.  We first note that because of the 
vector nature of the gauge propagator, there are no tadpole diagrams.
The diagram we consider then is the standard RPA which is a closed string of 
bubbles.  This is equivalent to calculating the free energy by integrating
out the spinon fields and obtaining an effective action for the gauge field.
The effective action is $S(a) = \sum_q \Pi(q) a^\dagger_q a_q$ with
$\Pi(q)$ from Eq. \ref{spinbubble}. Thus the partition
function is
\begin{equation}
Z = e^{-\beta F} = \int D\,a e^{-S(a)},
\end{equation}
where $F$ is the thermodynamic potential. Performing the functional integral,
we find
\begin{equation}
F = \frac{1}{\beta} \sum_{q_0} \int \frac{d\mathbf{q}}{(2\pi)^2}  \ln \Pi(q).
\end{equation}

We need to calculate the
susceptibility $\chi = - \frac{\partial^2 F}{\partial H^2}|_{H=0}$.
The spinon bubble in Eq. \ref{spinbubble} only depends on $H$
through the parameters $v^\sigma_\mathrm{F}$ and $k^\sigma_\mathrm{F}$
of $\Pi_\sigma(q)$.   Inserting Eq. \ref{dressedprop} into Eq. \ref{spinbubble},
we then Taylor expand to get
\begin{equation}
\Pi(q) = \Pi_0(q) + A(q) H^2 + O(H^3)
\end{equation}
where
\begin{equation}
\Pi_0(q) = \frac{ \left| q_0 \right| k_\mathrm{F}^2}
{\pi m \left(q_0 + \sqrt{q_0^2 + \frac{\mathbf{q}^2 k_\mathrm{F}^2}{m^2}} \right)}
+ \chi_\mathrm{d} \mathbf{q}^2
\end{equation}
and
\begin{equation}
A(q) = - \frac{\mathbf{q}^2 m^3 q_0 \sqrt{q_0^2+\frac{\mathbf{q}^2 k_\mathrm{F}^2}{m^2}}}
{2 \pi\left(\mathbf{q}^2 k_\mathrm{F}^2 + q_0^2 m^2\right)^2}.
\end{equation}
Plugging this expansion into the definition of the susceptibility, we find
that the correction to the zero field susceptibility due to this RPA diagram is 
\begin{equation}
\Delta \chi = - \frac{1}{\beta} \sum_{q_0} \int \frac{d\mathbf{q}}{(2\pi)^2}  \frac{ 2 A(q)}{\Pi_0(q)}.
\label{chiintegral}
\end{equation}
From the above functional forms of $A(q)$ and $\Pi_0(q)$, it is clear that $\Delta \chi > 0$.
Thus the RPA correction to the gauge field gives rise to an additional paramagnetic
contribution to the total susceptibility.  When the gauge field is gapped after
pairing, this correction is reduced.  This is consistent with the measured
drop that is seen experimentally.

An alternate way of calculating the spin susceptibility would be to dress
the spin-spin correlation function with the gauge field propagator
from Eq. \ref{dressedprop}.  From this one could calculate the susceptibility 
explicitly.  These two methods are the same because given the free energy
RPA diagrams, taking each derivative with respect to magnetic field is
equivalent to adding a spin-flip vertex.  By writing down all the topologically
inequivalent ways to add two spin-flip vertices, one generates the
diagrams of the spin-spin correlation function.
Kim \textit{et. al.} calculated the density-density correlation function
for small $\mathbf{q}$ and found that the singular portions coming the
self-energy and vertex corrections cancel and that there are only analytic 
corrections. \cite{Kim:94} This result applies equally to the spin-spin
correlation function. Our calculation corresponds to calculating
the numerical value of the nonsingular part of this correction.

\begin{figure}
\centerline{\includegraphics[width=2.5in]{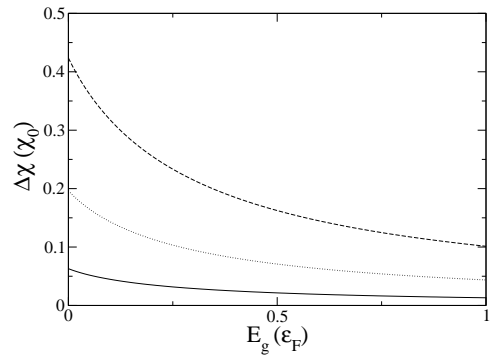}}
\caption{
$\Delta \chi$ in units of $\chi_0$ as a function of the pairing gap $E_g$ in
units of $\epsilon_\mathrm{F}$.
The solid, dotted, and dashed lines are for cutoffs $1$,$1.5$ and $2 k_\mathrm{F}$.
}
\label{chifig}
\end{figure}

We now proceed to calculate the numerical value of the shift in the
susceptibility in this RPA approximation for the gauge propagator. 
From the derived forms of $\Pi_0(q)$ and $A(q)$, we see that the integrand of Eq. 
\ref{chiintegral} increases as $q$ increases.  Thus to calculate a numerical value
we need to introduce a cutoff.  In the derivation of the polarization bubble
in Eq. \ref{dressedprop}, we assumed that $\mathbf{q} < k_\mathrm{F}$.  Thus we let
the energy integration go from zero to infinity and cutoff the momentum integral
at something of order $k_\mathrm{F}$.  In the standard calculation of the 
density-density polarization bubble for fermions in two-dimensions \cite{Stern:67}, 
one finds that the polarization bubbles dies off sharply as a square root for 
$q > 2 k_\mathrm{F}$ and is relatively flat inside the Fermi surface.  Thus in 
order to compare to experiment, we take the cutoff to be $2 k_\mathrm{F}$.
We write the paramagnetic correction at the RPA level in terms of the 
2-D Pauli paramagnetic susceptibility for a spin $1/2$ free fermion system, 
$\chi_0 = m / \pi$, where $m$ is the effective mass of the 
spinons.  For this cutoff, the extra paramagnetic contribution is 
$\Delta \chi \sim 0.42 \, \chi_0$.  This number, however, is strongly dependent on the 
cutoff as seen in Fig. \ref{chifig}.

We now introduce the gapping of the gauge field due to spinon pairing. Calling the gap 
energy $E_g$, once the spinons begin to pair, $\Pi(q)$ is shifted by this 
energy and as a result the denominator
of the integrand of Eq. \ref{chiintegral} becomes $\Pi_0(q) + E_g$.  As the
temperature decreases $E_g$ rises and the paramagnetic correction falls.
In Fig. \ref{chifig}, we plot $\Delta \chi(E_g)$, the extra paramagnetic
contribution due to these RPA diagrams, for different values of the cutoff.

\begin{figure}
\centerline{\includegraphics[width=2.5in]{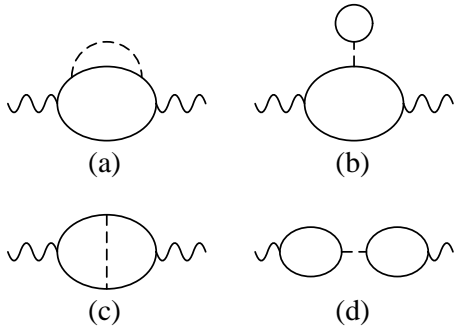}}
\caption{
Lowest order corrections in the short range interaction,
$U$, to the gauge propagator.
}
\label{Udiagram}
\end{figure}

We now consider the effect of the short range interaction on this calculation.  We
work perturbatively in $U$, the strength of this interaction.  We need to consider
the free energy diagrams that contain both the gauge field and the short range
interaction.  We continue to work using the RPA.  To lowest order in $U$ there
are four diagrams that contribute to the propagator by adding corrections
to the spinon bubble $\Pi(q)$.  These diagrams
are shown in Fig. \ref{Udiagram}.  
The top two diagrams are the standard Hartree and Fock corrections
to the spinon propagator.  These two are clearly related by an exchange.  One
can also see that the bottom two diagrams in Fig. \ref{Udiagram} are also related
to each other via an exchange.

First we consider the bottom two diagrams.  We denote the correction
to the total polarization for the vertex correction diagram 
(Fig. \ref{Udiagram}c) as $\Pi_\mathrm{c}$ 
and for its associated exchange diagram (Fig. \ref{Udiagram}d) as 
$\Pi_\mathrm{d}$.  In order to calculate the free energy, we close these diagrams with
the propagator calculated from the bare spinon bubble $D(q) = \Pi(q)^{-1}$. We find that
\begin{eqnarray}
\nonumber \Pi_\mathrm{c}(q) = \int dk\,dk'\, G(k) G(k+q) G(k') G(k'+q) \times \\
\frac{ \left[ 
\mathbf{k} \cdot \mathbf{k'} - (\mathbf{k} \cdot \mathbf{\hat{q}})
(\mathbf{k'} \cdot \mathbf{\hat{q}}) \right]}{m^2} U(k-k'),
\end{eqnarray}
\begin{eqnarray}
\nonumber \Pi_\mathrm{d}(q) = \int dk\,dk'\, G(k) G(k+q) G(k') G(k'+q) \times \\
\frac{ \left[ 
\mathbf{k} \cdot \mathbf{k'} - (\mathbf{k} \cdot \mathbf{\hat{q}})
(\mathbf{k'} \cdot \mathbf{\hat{q}}) \right]}{m^2}  U(q).
\end{eqnarray}
Assuming, for simplicity,
 that $U(q)$ is independent of $q$, it is clear that for both diagrams
the integral over $\theta_{\mathbf{k} \mathbf{q}}$, the angle between 
$\mathbf{k}$ and $\mathbf{q}$, is odd and thus gives zero.
This is because unlike in Eq. \ref{dressedprop} 
there is now only one power of $\mathbf{k}$ from the vertex.  We note that even without
the assumption of $U$ being momentum independent $\Pi_\mathrm{d}$ is zero,
but that the assumption is necessary for $\Pi_\mathrm{c}$.

The short range interaction can also dress the spinon propagator with 
Hartree-Fock corrections.  Instead of evaluating these
diagrams explicitly, we use the Dyson equation to calculate the dressed spinon
propagator for a given spin, $\tilde{G}_\sigma$.
The Hartree term, diagram Fig. \ref{Udiagram}b, gives a self energy $U n$ where $n$ is 
the total density of
spinons.  The Fock term, diagram Fig. \ref{Udiagram}a, gives a self energy 
$U n_\sigma$.  Thus we have
\begin{equation}
\tilde{G}_\sigma = \frac{1}{i\omega - \xi^{\sigma}_k - U n + U n_\sigma}.
\end{equation}
The total spinon density does not depend on magnetic fied, thus the Hartree term
contributes a field independent shift to the chemical potential which does not
effect the susceptibility and thus is dropped.  Applying a magnetic field 
does shift the density of a particular spin to
$n_\sigma(H) = n_\sigma(0) + \frac{dn}{d\mu} \delta \mu_\sigma$,
where $\delta \mu_\sigma = \pm 2 \mu_\mathrm{B} H$ for up and down
spins respectively and $\frac{dn}{d\mu} = \frac{m}{2\pi}$ is
the standard 2-D density of states. Thus,
\begin{equation}
\tilde{G}_\sigma = \frac{1}{i\omega - \left(
\epsilon_k - (\mu + \delta \mu_\sigma) - U(n_\sigma(0) + \frac{dn}{d\mu}
 \delta \mu_\sigma) \right)}.
\end{equation}
Again, dropping a shift in the chemical potential, this 
dressed spinon propagator
is the same as the undressed one except that when a magnetic field
$H$ is applied it responds to an effective field
$\tilde{H} = H \left( 1 + \frac{U m}{2 \pi} \right)$.

The gauge-field correction to the susceptibility 
including the short-range interaction to lowest order $\tilde{\chi}$
is thus
\begin{eqnarray}
\tilde{\chi} &=& -\frac{\partial^2 F}{\partial H^2}
= \left( \frac{\partial \tilde{H}}{\partial H} \right)^2 \chi \\
&=& \left( 1 + \frac{U m}{2 \pi} \right)^2 \chi,
\end{eqnarray}
where $\chi$ is the paramagnetic correction from Eq. \ref{chiintegral}.
Thus for $U>0$, we see that including the short
range interaction diagrams enhances the extra paramagnetic term
and thus leads to a larger drop in the susceptibility as the
pairing gap opens.

In conclusion, we have shown that the proposed Amperean pairing of
a spinon Fermi surface coupled to a $U(1)$ gauge field is consistent
with the experiments performed on the candidate material
$\kappa$-(BEDT-TTF)$_2$Cu$_2$(CN)$_3$.  In particular,  the
unconventional pairing of spinons on the same side of the Fermi surface
allows for a non-zero $T=0$ susceptibility despite the opening of a gap.
Also by calculating the effect of the gauge field on the paramagnetic
susceptibility, we found a drop in the susceptibility as the gauge field
becomes gapped due to pairing. This is consistent with the drop seen
in the experiments.  However, since the contribution to the 
susceptibility comes from the gauge field carrying large momentum,
the result is sensitive to the cut-off and the exact numerical
factor cannot be trusted.  Our goal is rather to show as a matter
of principle that there is a large paramagnetic contribution to the
susceptibility that is suppressed by the onset of pairing. 

We thank Y. Nakazawa and K. Kanoda for sharing their data prior to publication.
P.A.L acknowledges support by NSF grant DMR-0517222.


\appendix
\section{Derivation of $\Pi(q)$}

\begin{figure}
\centerline{\includegraphics[width=3in]{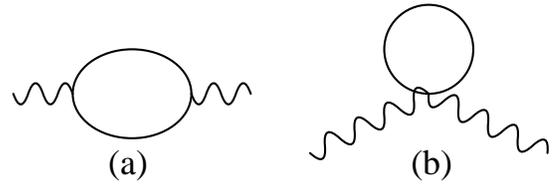}}
\caption{
The one loop corrections to the bare gauge propagator.
}
\label{dressedpi}
\end{figure}

In this appendix, we present the details of the calculation of the
gauge field propagator generated by the bare spinon bubble, Eq.
\ref{spinbubble}.  We define $\Pi(q) = \Pi_a(q) + \Pi_b(q)$, where 
the two one-loop diagrams that contribute are shown in Fig. \ref{dressedpi}.

We start with the diagram in Fig. \ref{dressedpi}a. Thus
\begin{equation}
\Pi_a(q) = \frac{2}{\beta} \sum_{k_0}
\int 
\frac{d\, \mathbf{k}}{(2\pi)^2} \, G\left(k+\frac{q}{2}\right) G\left(k-\frac{q}{2}\right)
\left| \frac{ \mathbf{k} \times \hat{\mathbf{q}}}{m} \right|^2,
\end{equation}
where the extra factor of two came from the spin summation.
Just as in the standard calculation of the polarization bubble we do the
Matsubara sum first and get
\begin{equation}
\Pi_a(q) = -2 \int \frac{d\,k d\,\theta}{(2\pi)^2} k 
\left(\frac{f(\xi_{\mathbf{k}+\frac{\mathbf{q}}{2}})-f(\xi_{\mathbf{k}-\frac{\mathbf{q}}{2}})} 
{i q_0 + \frac{\mathbf{k} \cdot \mathbf{q}}{m}} \right)
\frac{k^2 \sin^2(\theta)}{m^2},
\label{appendix1}
\end{equation}
where $\xi_k = \epsilon_k - \mu$, $\epsilon_k$ being the 
spinon dispersion relation and $\mu$ the Fermi energy.
If $|\mathbf{q}| < k_\mathrm{F}$ then the difference of the two
Fermi functions is only nonzero for a region of length 
$q \cos(\theta)$ around $k_\mathrm{F}$.  Assuming that q is much smaller than
$k_\mathrm{F}$, the k integration is just replaced by $k=k_\mathrm{F}$.
We now calculate the real and imaginary parts of $\Pi(q)$ and find 
\begin{eqnarray}
\Pi_a^1(q) &=& \frac{- k^3_\mathrm{F}}{2 \pi^2 m^2}
\int_0^{2\pi} \frac{v_\mathrm{F} q^2 \cos^2(\theta) \sin^2(\theta)}
{q^2_0+v^2_\mathrm{F} q^2 \cos^2(\theta)} \\
\Pi_a^2(q) &=& \frac{- k^3_\mathrm{F}}{2 \pi^2 m^2}
\int_0^{2\pi} \frac{- q_0 q \cos(\theta) \sin^2(\theta)}
{q^2_0+v^2_\mathrm{F} q^2 \cos^2(\theta)}.
\end{eqnarray}
Doing these integrals, we find that the imaginary part is zero and that
the real part gives,
\begin{equation}
\Pi_a(q) = \frac{-k^2_\mathrm{F}}{2 \pi m} - \frac{m}{2 \pi} \frac{1}{\mathbf{q}^2}
\left[2 q^2_0 - 2 \left|q_0\right| \sqrt{v^2_\mathrm{F} \mathbf{q}^2 + q^2_0}\right].
\label{EQ:pi1}
\end{equation}
This gives the dominate correction for small $\mathbf{q}$ for the non-static
polarization bubble.
For the static part of the polarization bubble, $q_0=0$, this 
$\frac{1}{\mathbf{q}^2}$ term vanishes, so we have to relax the approximation
to the difference of the Fermi functions in order to get
the $\mathbf{q}$ dependence in the static case.  We now write
\begin{equation}
\xi_{\mathbf{k} \pm \frac{\mathbf{q}}{2}} = \xi_\mathbf{k} + \Delta_\pm,
\end{equation}
where $\Delta_\pm = \pm \frac{\mathbf{k} \cdot \mathbf{q}}{2 m} +
\frac{\mathbf{q}^2}{8 m}$.
We Taylor expand each distribution function in $\Delta_\pm$.  We work
to third order in $\Delta_\pm$ in order to generate all the terms up to order $q^2$.
Dropping all higher powers of $q$, we find that in the static limit, $q_0=0$,
\begin{eqnarray}
\nonumber
\Pi_a(q) &=& \frac{-1}{2 \pi^2}\int d^2k \frac{k^2 \sin^2(\theta)}{m^2} \times \\
& &\left[ n'(\xi_k) + \frac{q^2}{8 m} n''(\xi_k) + 
\frac{k^2 q^2}{96 m^2} n'''(\xi_k) \right].
\end{eqnarray}
Which gives
\begin{equation}
\Pi_a(q) = \frac{-1}{\pi} \left[ \frac{k_\mathrm{F}^2}{2 m} - \frac{q^2}{12 m}
\right].
\label{EQ:staticpart}
\end{equation}

We now compute the effect of the diagram shown in Fig. \ref{dressedpi}b.
The spinon loop here just gives a density.  So that this diagram evaluates to
\begin{equation}
\Pi_b(q) = \frac{-2}{m} (-n) = \frac{k^2_\mathrm{F}}{2 \pi m}
\label{EQ:piloop}
\end{equation}

Putting Eq. \ref{EQ:pi1}, Eq. \ref{EQ:staticpart} and Eq. \ref{EQ:piloop} 
together, we get that
\begin{equation}
\Pi(q) = \frac{\gamma v_\mathrm{F} \left|q_0\right|}{\sqrt{v^2_\mathrm{F} \mathbf{q}^2 +q^2_0}
+ q_0 } + \chi_\mathrm{d} \mathbf{q}^2
\end{equation}
where $\gamma = \frac{ k_\mathrm{F}}{m}$ and $\chi_\mathrm{d} = \frac{1}{12 m \pi}$.

Taking the limit $q_0 \ll v_\mathrm{F} q$, we recover the result from earlier
papers $\Pi(q) = \gamma \frac{\left| q_0 \right|}{\left| q \right|} + 
\chi_\mathrm{d} q^2$.
\cite{Lee:92,Ioffe:89}
Note that the actually value of $\chi_\mathrm{d}$ differs from \cite{Lee:92}
by a factor of two because in that paper, there is a factor of two error in the
form of the 2-D Landau diamagnetic susceptibility that gives rise to 
$\chi_\mathrm{d}$.

\bibliography{tjbib}

\end{document}